\documentclass[12pt,preprint]{aastex}

\usepackage{emulateapj5}
\usepackage{graphics,graphicx}
\usepackage{fancyheadings}
\usepackage{color}
\usepackage{natbib}

\newcommand{\aox}{\ifmmode \alpha_{ox}\else$\alpha_{ox}$\fi}
\newcommand{\mone}{\ifmmode ^{-1}\else$^{-1}$\fi}
\newcommand{\mtwo}{\ifmmode ^{-2}\else$^{-2}$\fi}
\newcommand{\degs}{\ifmmode ^{\circ}\else$^{\circ}$\fi}
\newcommand{\mv}{\ifmmode {m_{V}}\else${m_{V}}$\fi}

\newcommand{\gae}{\mathrel{\raise .4ex\hbox{\rlap{$>$}\lower 1.2ex\hbox{$\sim$}
} }}

\shorttitle{3C186 X-ray cluster}

\slugcomment{\today}

\begin{document}


\title{High redshift X-ray cooling-core cluster associated with the luminous radio
  loud quasar 3C186}

\author{Aneta Siemiginowska$^{1}$, D.J. Burke$^{1}$, Thomas~L.~Aldcroft$^{1}$, D.M.~Worrall$^2$, S.Allen$^3$, Jill Bechtold$^4$, Tracy Clarke$^5$, C.C.Cheung$^6$}

\affil{$^{1}$ Harvard Smithsonian Center for Astrophysics, 60 Garden St, 
Cambridge, MA 02138}

\affil{$^2$ H.H. Wills Physics Laboratory, University of Bristol,
  Tyndall Avenue, Bristol BS8 1TL, UK}

\affil{$^3$ Kavli Institute for Particle Astrophysics and Cosmology,
  Stanford University, 382 Via Pueblo Mall, Stanford, CA 94305-4060,
  USA}

\affil{$^4$ Steward Observatory, University of Arizona, Tucson, AZ}

\affil{$^5$ Naval Research Laboratory, Code 7200, 4555 Overlook Ave SW, Washington, 
DC 20375}
 
\affil{$^6$ National Research Council Research Associate, Space
Science Division, Naval Research Laboratory, Washington, DC 20375,
USA}

\email{asiemiginowska@cfa.harvard.edu}


\label{firstpage}

\begin{abstract}

  We present the first results from a new, deep (200ks) {\it Chandra}
  observation of the X-ray luminous galaxy cluster surrounding the
  powerful ($L\sim10^{47}$\,erg~s$^{-1}$), high-redshift ($z=1.067$),
  compact-steep-spectrum radio-loud quasar 3C\,186. The diffuse X-ray
  emission from the cluster has a roughly ellipsoidal shape and
  extends out to radii of at least $\sim 60$~arcsec ($\sim 500$\,kpc).
  The centroid of the diffuse X-ray emission is offset by 0.68$\pm0.11
  \arcsec$ ($\sim5.5\pm0.9$\,kpc) from the position of the quasar. We
  measure a cluster mass within the radius at which the mean enclosed
  density is 2500 times the critical density, $r_{2500}=
  283^{+18}_{-13}$\,kpc, of
  1.02$^{+0.21}_{-0.14}\times10^{14}$M$_{\odot}$. The gas mass
  fraction within this radius is $f_{gas}=0.129^{+0.015}_{-0.016}$.
  This value is consistent with measurements at lower redshifts and
  implies minimal evolution in the $f_{gas}(z)$ relation for hot,
  massive clusters at $0<z<1.1$.  The measured metal abundance of
  0.42$^{+0.08}_{-0.07}$ Solar is consistent with the abundance
  observed in other massive, high redshift clusters.  The
  spatially-resolved temperature profile for the cluster shows a drop
  in temperature, from $kT \sim 8$\,keV to $kT \sim 3$\,keV, in its
  central regions that is characteristic of cooling core clusters.
  This is the first spectroscopic identification of a cooling core
  cluster at $z>1$. We measure cooling times for the X-ray emitting
  gas at radii of 50\,kpc and 25\,kpc of $1.7\pm0.2 \times 10^9$~years
  and $7.5\pm2.6 \times 10^8$~years, as well as a nominal cooling rate
  (in the absence of heating) of $400\pm190\, \rm
  M_{\odot}$year$^{-1}$ within the central 100\,kpc. In principle, the
  cooling gas can supply enough fuel to support the growth of the
  supermassive black hole and to power the luminous quasar. The
  radiative power of the quasar exceeds by a factor of 10 the
  kinematic power of the central radio source, suggesting that
  radiative heating may be important at intermittent intervals in
  cluster cores.

\end{abstract}


\section{Introduction}
\label{sec:intro}

Recent X-ray observations of nearby galaxy clusters show that powerful
outbursts of their central cD galaxies have imprinted a rich variety
of structures onto the X-ray emitting gas (see \citet{mcnamara2007}
for a review). The amount of energy supplied into the cluster gas by
these outbursts can prevent the clusters from cooling
\citep{mcnamara2005}.  The average power released in outbursts exceeds
$\sim 10^{45}$~erg~s$^{-1}$. This is equivalent to the typical
radiative power of a quasar. However, the central cD galaxy is usually
observed in a quiescent, non-luminous, state. The impact of a luminous
quasar on the cluster gas has not been widely explored so far.

On the other hand, it has been known for a long time that powerful
radio-loud quasars are associated with rich galaxy environments
\citep{smith1990,ellingson1991,yee1993}, and so we should be able to
find luminous radio loud quasars in X-ray-bright clusters. Much
remains unknown about the details of how quasar activity is triggered,
the way in which the accretion proceeds, the impact of the quasar on
its galactic and cluster environment, and which physical processes are
important as a function of redshift. X-ray clusters retain the history
of the cD activity and studies of quasars in X-ray clusters can
provide some answers to these questions.

There have been a few successful searches for the diffuse X-ray
cluster emission surrounding powerful quasars and radio galaxies using
ROSAT \citep{cf1993, worrall94, hall1995, hall1997, Crawford:1999,
  hardcastle1999, sarazin1999, worrall2000}.  \cite{hall1997}
discussed the ROSAT X-ray data for five quasars located in X-ray
clusters, and concluded that simple models for triggering the quasars
(e.g., cooling flows, a low-pressure ISM, or low velocity dispersions)
do not work, suggesting instead that mergers and strong interactions
are critical in delivering the gas to the quasar host.
\cite{hardcastle1999} analyzed a large sample of powerful radio
sources and found cases of extended X-ray emission that are likely to
be of cluster origin. However, the number of such cases was limited to
eight by ROSAT's sensitivity and some of these were not confirmed by
{\it Chandra} observations \citep{cf2003,worrall2004}. Extended X-ray
emission can be associated alternatively with radio structures such as
large scale jets, knots, hot spots, and lobes
\citep[e.g.,][]{celotti2004,kataoka2005,croston2005,harris2006} and
disentangling the cluster emission from the other X-ray emission
components has been challenging.  The {\it Chandra\/} X-ray
Observatory\citep{weiss2000} has both the spatial resolution and the
dynamic range necessary to study diffuse X-ray emission in the
vicinity of a strong point source, and allows the separation of
cluster X-ray emission from that of jets and other structures.

\cite{belsole2007} studied a {\it Chandra\/} and XMM-{\it Newton}
sample of twenty powerful radio sources at $z>0.5$, including two
core-dominated quasars, and found diffuse X-ray emission in 60\% of
the sample. The diffuse emission was faint, but the luminosities were
consistent with non-quasar-host X-ray clusters at similar and lower
redshifts, and the work found no difference between the cluster
environments of quasars and radio galaxies.

We discovered a bright X-ray cluster in a pointed {\it Chandra\/}
observation of the radio-loud compact steep spectrum (CSS) quasar 3C\,186
at the redshift of $ z=1.067$
\citep{siem2005}. This observation, although only $\sim30$~ksec
long, provided an X-ray luminosity measurement, $L_{(0.5-2 keV)} = 6
\times10^{44}$~erg~s$^{-1}$, a cluster temperature ($kT =
5.2^{+1.3}_{-0.9}$~keV) and a gas-mass fraction 
($f_{\rm gas}(r_{2500}) \sim 0.13\pm0.08$) that were
typical of other massive, relaxed clusters
\citep{alexy2002,allen2008}. 
The 3C\,186 X-ray cluster is more luminous than
the
\cite{belsole2007} clusters, and provides a unique opportunity
to study a luminous cluster associated with a quasar at high
redshift.  The powerful and luminous quasar, $L_{bol}\sim10^{47}$
erg~s$^{-1}$, is located well within the diffuse X-ray emission.

Here we report the results from a 200~ks deep follow-up {\it
  Chandra\/} observation of 3C\,186, made in order to study the quasar
and its associated cluster in greater detail.  Relatively few massive,
relaxed X-ray clusters at $z>1$ are known.  In addition, the 3C\,186
cluster is one of very few X-ray clusters that is both associated with
a quasar and bright enough for detailed study.  We give details of the
{\it Chandra\/} observation in Section 2, describe image and spectral
analyzes in Section 3, and present a discussion of the results in
Section~4.

Throughout this paper we use cosmological parameters based on
WMAP measurements \citep{spergel2003}:
$H_0=$71~km~s$^{-1}$~Mpc$^{-1}$, $\Omega_M = 0.27$, and $\Omega_{\rm
vac} = 0.73$. At $z= 1.067$, 1\arcsec\ corresponds to $\sim$8.163~kpc.

\section{Chandra Observations}
\label{sec:data}


\begin{table*}
{\scriptsize
\noindent
\caption[]{\label{table-1} 
{\it Chandra} Observations}
\begin{center}
\begin{tabular}{cccccrc}
\hline\hline
\\
OBSID & Exposure (ks) & Date & CCD & Inner$^a$ 2.5-6.0'' & Outer$^a$ 6-20''\\
\\
\hline
\\
9407 & 66.3  & 2007-12-03 & ACIS-23567 & 500.0$\pm22.8$ & 1000.7$\pm34.8$ & \\
9774 & 75.1  & 2007-12-06 & ACIS-23567 & 613.0$\pm25.3$ & 1110.6$\pm36.9$ & \\
9775 & 15.9  & 2007-12-08 & ACIS-23567 & 114.8$\pm11.0$ & 241.5$\pm17.4 $ &\\
9408 & 39.6  & 2007-12-11 & ACIS-23567 & 306.9$\pm17.8$ & 627.9$\pm27.3$ &\\
\\
\hline\hline
\end{tabular}

$^a$ net counts within the energy range 0.3-7~keV in annuli
centered on the quasar whose radii are given in arcsec.
\end{center}
}
\end{table*}

3C\,186 was observed with the {\it Chandra\/} ACIS-S CCD in December
2007. Due to scheduling constraints the observation was split into
four separate pointings that sum to a total exposure of 197~ks (see
Table~\ref{table-1}). The quasar was placed on the back-illuminated
CCD (S3) and was offset by -1 arcmin in Y coordinates to make sure
that the cluster is not affected by a chip gap. The observation was
made in VFAINT mode and full-window mode.  Table~\ref{table-1} shows
which CCD chips were active.  The observation was not affected by
solar flares, and the background was quiet for the entire observation.
The quasar is slightly offset in obsid 9408 with respect to the other
three observations in Z coordinates. However, the four observations
were performed with similar enough configurations that they could be
merged together for the purpose of image analysis. Note that the
exposure maps are flat on the scale size of the cluster in all four
observations.  Analysis was performed with the CIAO version 4 software
using CALDB version 4.2. All modeling was done in {\it Sherpa}
\citep{freeman2001,refsdal2009}. We used the Cash and Cstat fitting
statistics \citep{cash1979} and the Nelder-Mead optimization method
\citep{neldermead1965}. We did not subtract the background, but
included a background contribution in the model expressions for all of
our model fitting.

The events have been filtered to remove the VFAINT background
events. The standard $\pm 0.5$ pixel randomization has also been
removed before merging the four observations. The ACIS image of the
merged observation is shown in Figure~\ref{acis}. The bright quasar is
located within diffuse X-ray emission that is visible on a scale
exceeding $\sim$30\arcsec\,(250~kpc).

\section{Data Analysis and Results}
\label{sec:results}

\subsection{Image Analysis}

\begin{figure*}
\begin{center}
\includegraphics[width=5in]{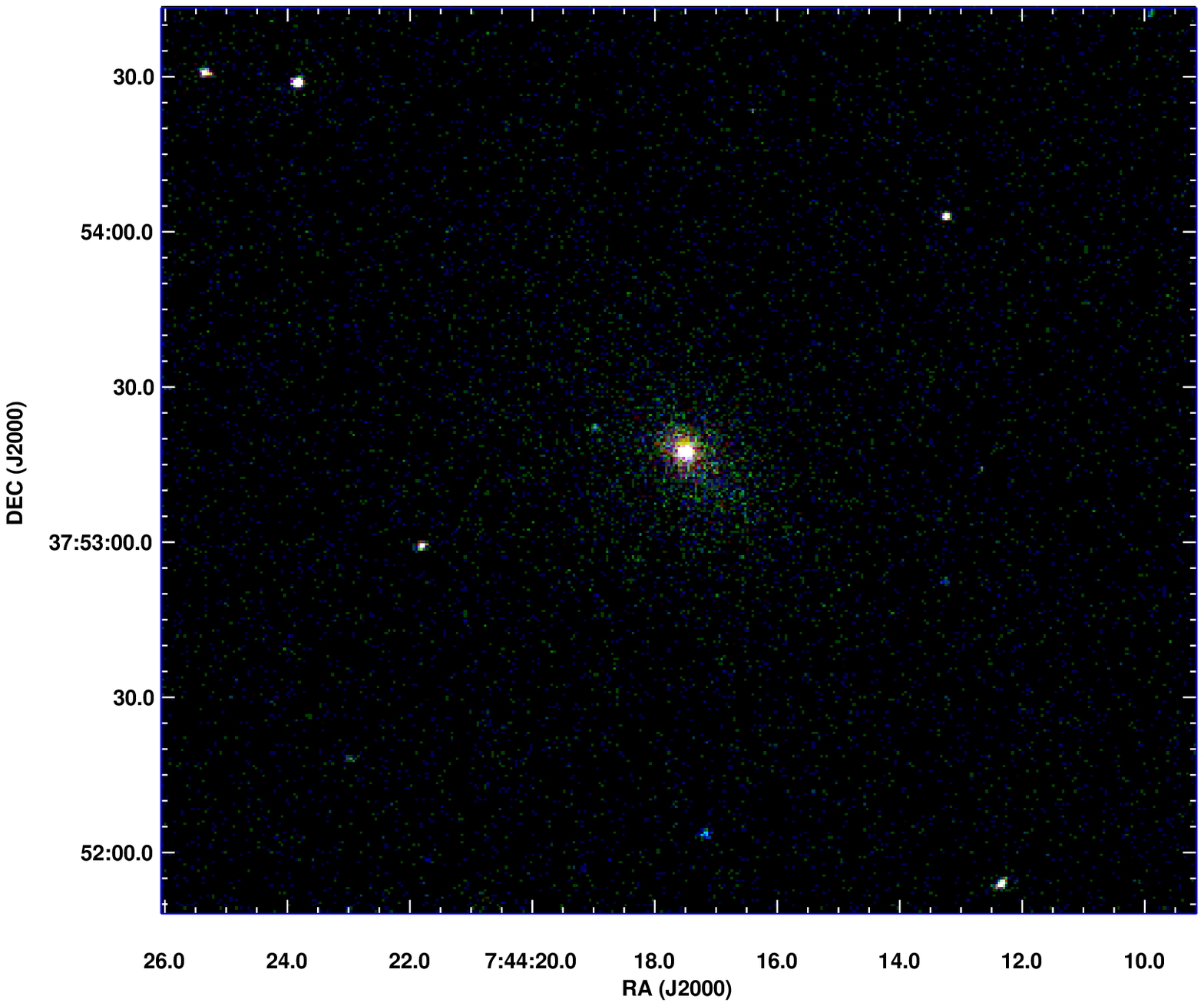}
\includegraphics[width=5in]{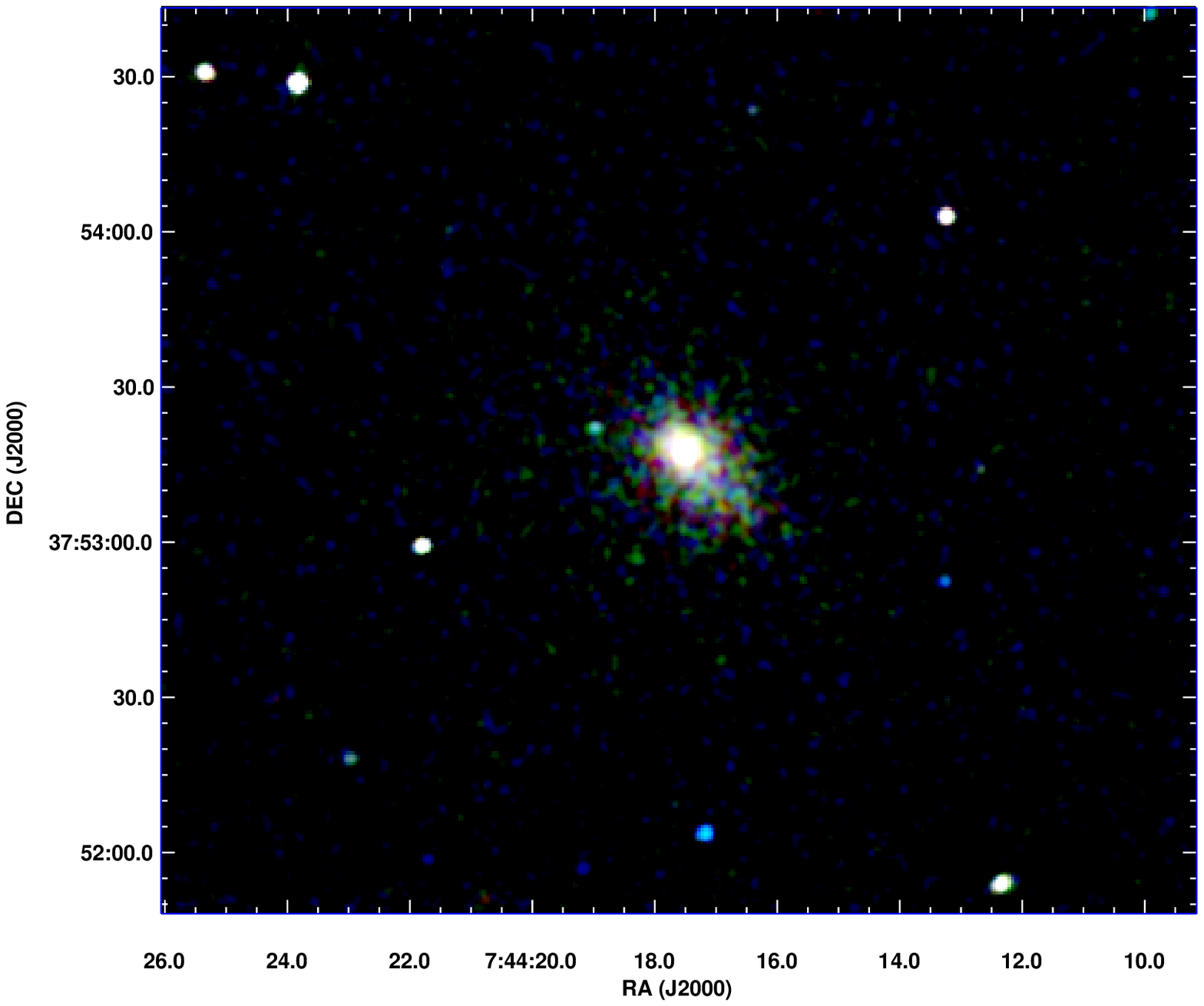}
\caption{\small RGB color {\it Chandra\/} ACIS-S images of the 3C\,186
  X-ray cluster.  The four individual observations have been merged
  into one image. Colors represent different energies:
  red:0.3-1.5~keV, green:1.5-2.5~keV and blue:2.5-7~keV.  {\bf Top:}
  Images binned to ACIS-S pixels with the standard size of
  1~pixel=0.492 arcsec.  {\bf Bottom:} The image has been smoothed by a
  Gaussian function of $\sigma=2.46\arcsec$. 
}
\end{center}
\label{acis}
\end{figure*}

We performed 2D image analysis using the merged data to evaluate the
spatial extent of the diffuse emission, and the quasar contribution to
the diffuse emission within $<$5~arcsec of the core.
Figure~\ref{acis} shows the merged ACIS-S images in three RGB colors
representing different energy ranges: red corresponds to 0.3-1.5~keV,
green to 1.5-2.5~keV, and blue to 2.5-7~keV.  We show a binned image
with the standard ACIS-S pixel size of 0.492$\arcsec$ and an image
that is smoothed with a Gaussian kernel ($\sigma=2.46\arcsec$). The
cluster emission is relatively smooth, elliptical, and elongated in
NE-SW direction. We measure (using ds9) the largest extent of the
emission to be $\sim 128\arcsec$ along the major axis at PA = 43~deg,
i.e. $\sim 60 \arcsec$ ($> 500$~kpc) distance from the quasar. We
selected the events within a 0.5-7~keV energy range for the spatial
modeling.

Effects of the strong quasar emission have been taken into account in
the analysis of the cluster data.  Note that point sources in the field
of view in Figure~\ref{acis} indicate the size of the PSF for a source
of typical strength.  We
adopted the observed quasar spectrum (a power law with a photon index of
$\Gamma=1.9$) as the input model for
CHART\footnote{http://cxc.harvard.edu/chart/index.html} simulations of
the quasar's PSF, and created a very high signal-to-noise PSF model image for
use in our analysis.

In our two-dimensional analysis we fitted both the cluster centroid
and the quasar position to determine whether they are co-aligned. We
tried circular models and also allowed the cluster gas to have an
ellipticity. We used {\it Sherpa} and adopted a 2D
Gaussian\footnote{We used Gaussian model instead of a Delta function
  to account for an unknown aspect ``blur'' in MARX.}  model for the
quasar and a 2D Beta model for the cluster. We also included a
constant model to account for the background.  Because the exposure
map is uniform across the cluster region we did not include the
exposure map in this analysis, and so we worked with the count data
and used Poisson statistics in fitting.  However, we have confirmed
that including the exposure map gives consistent parameters with the
ones reported here.

We used {\it Sherpa\/} to convolve the 2D model with the simulated 2D
PSF, and fitted the result to the data. The models and best-fit
parameters are given in Table~\ref{table-2a}.

The 2D fit finds a cluster core radius of
$r_{core}=3.51\pm0.31$~arcsec ($\sim$28.6$\pm2.5$~kpc) and
$\beta=0.48\pm0.17$ for elliptical models and
$r_{core}=3.06\pm0.25$~arcsec ($\sim 25.0 \pm 2.0$~kpc) and
$\beta=0.48\pm0.17$ for the circular models. 
We note that there are systematic uncertainties present in the 2D
analysis.  The circular and elliptical fits have small statistical
errors, but the difference between the parameters obtained in circular
and elliptical models, in particular the cluster core radius, could be
larger than the 1$\sigma$ statistical errors reported in
Table~\ref{table-2a}.

\begin{figure*}
\includegraphics[width=\columnwidth]{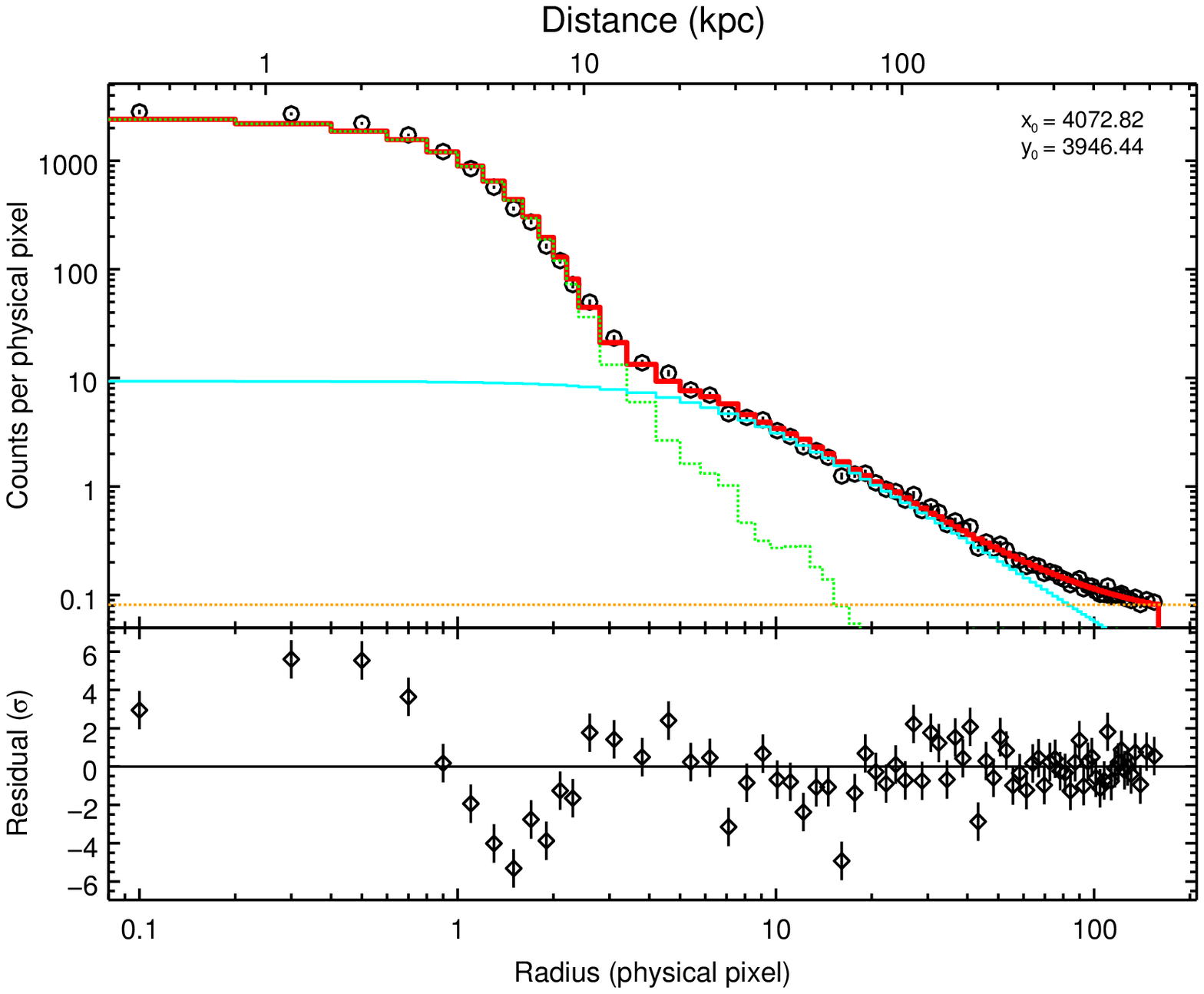}
\caption{\small {\bf Top panel:} Surface brightness profiles (in units
  of counts per pixel) centered on the quasar extracted from the data
  and the 2D circular models (see Table 2 for the best fit
  parameters).  The location of the quasar centroid in physical
  coordinates is marked in the upper right corner. The surface
  brightness profile from the ACIS-S data is shown as black points
  with errorbars: data is binned to ensure at least 200 counts per
  bin. The model profiles are marked with solid lines: red - full
  model (qso + clus+ bgnd as defined in Table 2) convolved with ChART
  PSF; green - quasar (gauss2d only) model convolved with PSF (ie no
  cluster or background component).  blue - cluster (beta2d), brown -
  background (cont2d).  {\bf Bottom panel: } Residuals between the
  full model and the data in units of $\sigma$.  Physical pixel size
  is equal to 0.492$\arcsec$, the size of the ACIS-S pixel. }
\label{profile}
\end{figure*}

The best-fit locations of the quasar and the cluster centroids are
offset by 0.68~arcsec, which is significantly greater than the
corresponding uncertainty of 0.11~arcsec.  We centered the surface
brightness profile on the peak of the X-ray emission, while the true
cluster centroid is slightly offset from the peak.
Figure~\ref{profile} shows surface-brightness profiles centered on the
quasar and obtained from the data and the 2D best-fit circular models.
The quasar emission dominates over the cluster emission within about
1.5 arcsec of the core and it is unresolved. At $2.5\arcsec$ distance
from the peak the quasar contributes only about 10\% of the total
counts. The X-ray cluster emission dominates outside that region and
exceeds the background level up to $\sim 40 \arcsec$ ($\sim 320$~kpc)
from the quasar core. The cluster is detected at 3$\sigma$ to $\sim
285$~kpc.

\begin{table*}
\label{tab2}
{\scriptsize
\noindent
\caption[]{\label{table-2a} 
Best Fit 2D Model Parameters$^a$}
\begin{center}
\begin{tabular}{llccccccc}
\hline\hline
\\
   Parameter  &  Circular Models$^b$ &   Elliptical Models $^b$    &  Units \\
\\
\hline
\\
qso.fwhm & 0.43${^+_-0.02}$      & 0.26  ${^+_-0.02} $ &   arcsec\\
qso.xpos & 4072.82${^+_- 0.01}$ & 4072.82 ${^+_- 0.01}$ & physical\\
qso.ypos & 3946.44${^+_- 0.01} $ & 3946.43 ${^+_- 0.01} $ & physical\\
qso.ellip & -- 		& 0.115 ${^+_- 0.097}$ & \\
qso.theta & --          & 1.82 ${^+_-0.45} $ & radians\\
qso.ampl & 2018.5$^{+165.8}_{-148.6}$ & 1497.1 ${^+_- 138.6} $ & counts\\
clus.r0 & 3.03${^+_-0.25} $  & 3.58 ${^+_-0.31} $ & arcsec\\
clus.xpos & 4072.98${^+_-0.21}$ & 4072.96 ${^+_- 0.21} $ &\\
clus.ypos & 3945.05${^+_-0.21}$ & 3945.06 ${^+_- 0.22} $& \\
clus.ellip & -- & 0.300 ${^+_- 0.015} $ &\\
clus.theta & -- & 2.41 ${^+_- 0.03} $& radians\\
clus.ampl &  0.42${^+_- 0.04}$ & 0.44 ${^+_- 0.04} $ & counts\\
clus.beta & 0.48 ${^+_-0.17}$ & 0.48 ${^+_-0.17} $ & \\
bgnd.c0 & 0.0023${^+_-0.0002}$  & 0.0024 ${^+_-0.0001} $ & counts\\
\hline\hline
\end{tabular}

\smallskip

$^a$ Fit was performed on non-background subtracted image and the background
model was included as a part of the model component: 
qso - 2D gaussian model; bgnd - 2D constant background model
\\
 clus - 2D beta model: $f(x,y) = f(r) =A (1+({r \over r_0})^2)^{-3\beta+0.5}$;\\
$r(x,y)=\sqrt {x\prime ^2(1-\epsilon)^2 + y\prime^2}/(1-\epsilon)$; \\
$x\prime = (x-x_o)\cos\theta + (y-y_o)\sin\theta, 
y\prime = (y-y_o)\cos\theta - (x-x_o)\sin\theta
$
$\theta$ - the angle of ellipticity, $\epsilon$ - ellipticity.

$^b$ 1$\sigma$ uncertainties are shown for one interesting parameter.
\end{center}
}
\end{table*}

\medskip


\subsection{Spectral Modeling of the Cluster Emission}
\label{sec:spectra}

\begin{figure*}
\includegraphics[width=\columnwidth]{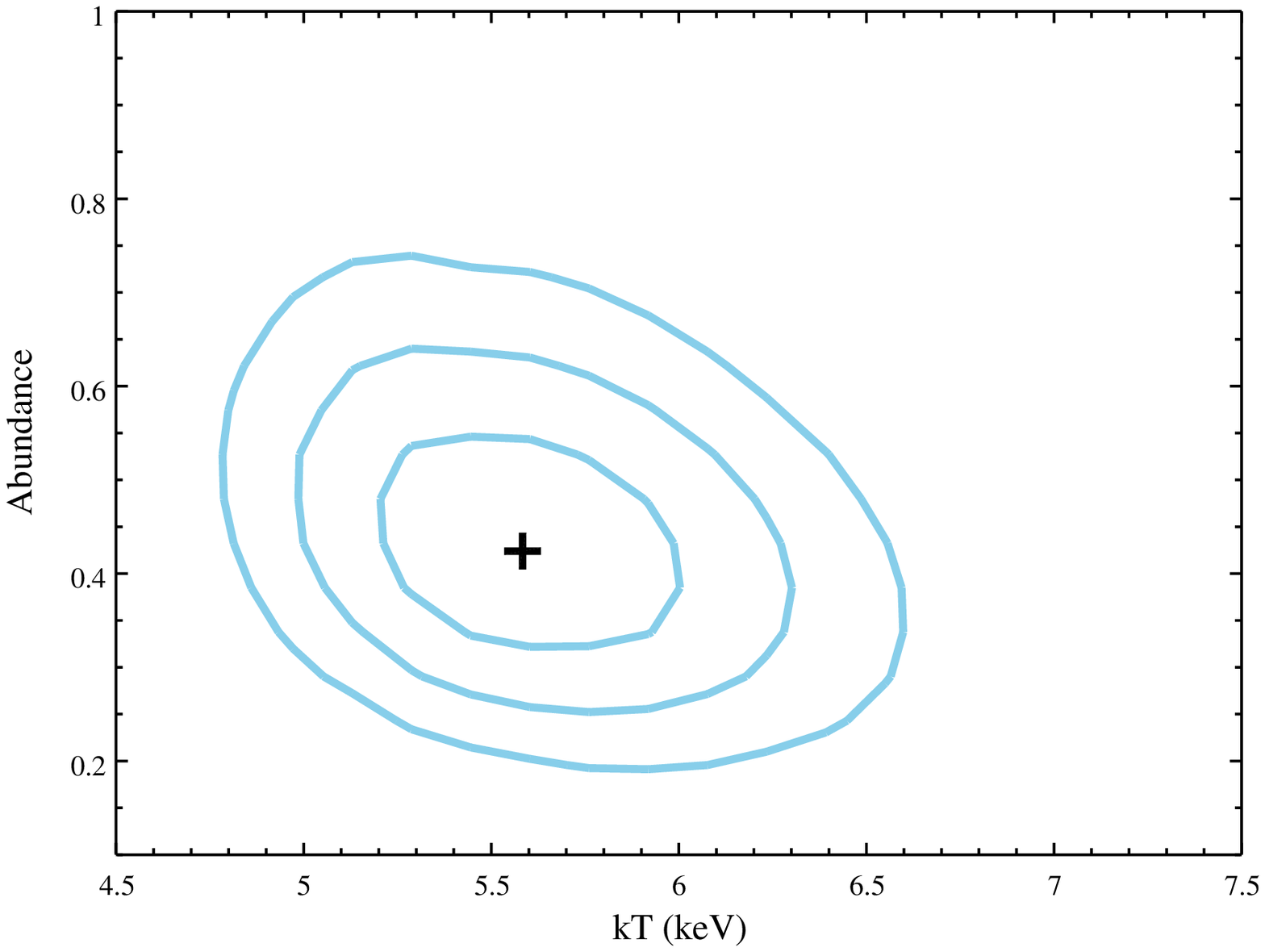}
\caption{\small The confidence levels for the temperature and
  abundance parameters obtained in a global model fit to the cluster
  described in Sec.~\ref{sec:spectra}. The location of the best fit
  parameters is marked by a cross.  Three levels 1$\sigma$, 2$\sigma$
  3$\sigma$ contours are shown.}
\label{levels}
\end{figure*}

Based on our 2D image modeling results we assume that the X-ray
emission outside the central 2.75$\arcsec$ ($\sim 22$~kpc) radius
circular region is dominated by cluster emission (see analysis of
  the quasar contribution to the observed emission below). Note that
3C~186 is a compact radio source with a linear size of 2$\arcsec$, so
any X-ray emission associated with the radio source would be contained
within this central region.  We performed spectral analysis using each
individual observation (obsid) and the corresponding calibration
files. To obtain a global temperature for the entire cluster we
extracted the spectrum from each event file (4 obsids) assuming a
large elliptical region with the semi-minor and semi-major axis of
20$\arcsec$ and 30$\arcsec$ at PA~315 degree and ignoring the inner
circle with radius of 2.75~arcsec dominated by the quasar.  The
background files were extracted from regions on the same CCD located
outside the source region excluding all the detected point sources.
The APEC thermal plasma model at redshift $z=1.067$ was fitted to the
four spectra simultaneously, giving a best global temperature of
kT=5.58$^{+0.28}_{-0.27}$~keV, a metallicity of 0.42$^{+0.08}_{-0.07}$
Solar and a soft-band flux F$_{0.5-2} =
8.27\times10^{-14}$~erg~sec$^{-1}$~cm$^{-2}$ (Fig.~\ref{levels}). The
best fit global cluster temperature is in agreement with the single
temperature measured in the first short observation \citep{siem2005}.


In order to look for any temperature gradient we extracted the spectra
from each event file (4 obsid) assuming seven circular annuli centered
on the quasar. The annuli cover the range from 2.75$\arcsec$ to
30$\arcsec$. We list the angular ranges spanned by the annuli in
Table~\ref{table-2b}.  All spectra were taken from individual
observations to properly account for the instrumental effects. They
were simultaneously fitted using all available counts for each annuli
in the energy range 0.5-7~keV. We accounted for any background
contribution using a complex empirical model (a combination of an 8th
order polynomial and five Gaussian lines) that was first fitted to the
ACIS-S ``blank-sky'' background data (see Appendix A1 for more
details). We next fit this background model to the background spectra
from each observation to check how well the model describes our data.
In the simultaneous fit of the source and background model to the
cluster spectra we varied only the background model normalization and
kept all the other background model parameters frozen. We note that
this is reasonable, as the fraction of background counts is typically
lower than a few percent and exceeds 23\% only in the outermost
annulus.

\begin{figure*}
\includegraphics[width=\columnwidth]{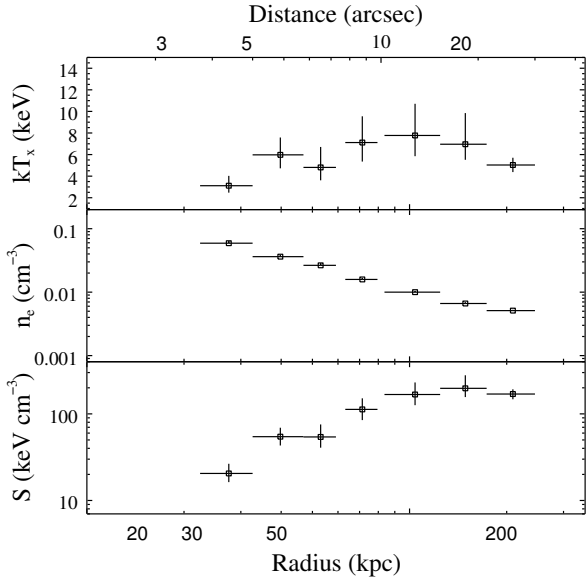}
\caption{\small Deprojected temperature, density and entropy profiles
  for the best fit model parameters shown in Table~\ref{table-2b}. }
\label{temp}
\end{figure*}

Table~\ref{table-2b} shows the best fit parameters for the applied
model. We assumed an APEC model, included a correction for the
Galactic absorption with an equivalent Hydrogen column density of $N_H
=5.64\times 10^{20}$~cm$^{-2}$, and assumed cluster metal abundances
of 0.3 in respect to Solar.  We developed a {\tt deproject}
model\footnote{http://cxc.harvard.edu/contrib/deproject/} in {\it
  Sherpa} based on the description in \citet{fabian1981} and
\citet{kriss1983}. We assume a spherical geometry and the radial sizes
of the shells given by each individual annulus. The fit starts at the
outermost annulus and proceeds towards the innermost one taking into
account the contributions from outer annuli to the fitting of the
inner one. The best-fit model gives the deprojected temperatures,
normalizations, and densities listed in Table~\ref{table-2b}. We plot
deprojected temperature, density and entropy profiles in
Figure~\ref{temp}.  These profiles are consistent with the cluster
having a cooling core.

\begin{figure*}
\includegraphics[width=\columnwidth]{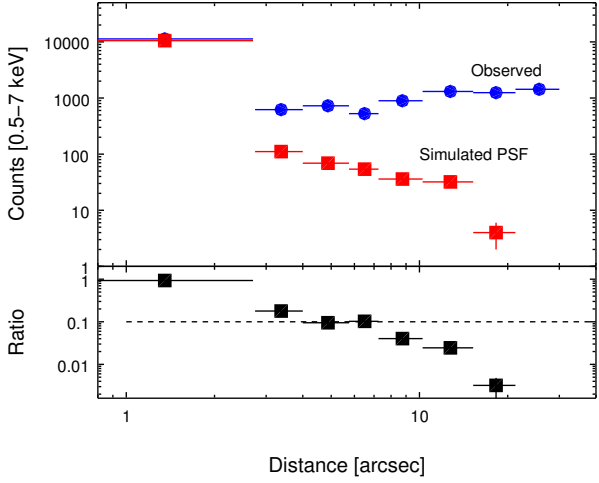}
\caption{\small {\bf Top panel:} A number of counts in the spectral
  extraction regions for the observed (quasar+cluster) emission (blue
  circles) and the CHART simulated PSF - red squares. The first data
  point shows the PSF normalized to the observed counts in the
  circular region with r=2.7$\arcsec$ centered on the quasar. {\bf Lower
    panel:} The ratio of the simulated to the observed counts in each
  spectral region. The dashed line marks 0.1 value.}
\label{fractions}
\end{figure*}

We followed \cite{russell2009}, using the simulated {\it Chandra} PSF
to understand the contribution from the central quasar to each
individual annulus and the effects associated with a possible
contamination of the cluster spectrum. We used CHART to simulate the
quasar (point source) photons scattered by the {\it Chandra} mirrors,
and MARX to project them onto the ACIS-S detector at the exact
pointing as in the {\it Chandra} observations of 3C~186. In the CHART
simulations we assumed the quasar photon flux to be described by the
best model parameters fit to the spectrum extracted from a $r=1.75
\arcsec$ circular region centered on the quasar, i.e.  an absorbed
power law with $N_H$ and $\Gamma = 1.9$.  The simulated quasar image
reflects both the scattering of the {\it Chandra} mirrors and an
additional ``blur'' (parameter set to 0.3 in MARX) due to the standard
dither and the aspect uncertainty. In Figure~\ref{fractions} we show a
number of counts from the simulated quasar image in comparison to a
number of counts detected by {\it Chandra} in each annulus assumed for
the spectral modeling.  The fraction of the total observed counts that
can be associated with the quasar is shown in the bottom panel of the
figure. It exceeds 10\% only in the first innermost annulus.

In order to check the effects of quasar contamination on the spectral
results, additional fits were performed wherein we included
appropriately normalized power law model components to the cluster
model for each annulus (see Table~\ref{table-4}).  The resulting
temperatures and normalizations are in excellent agreement with the
deprojected model results shown in Table 3.  The only noticeable
difference is for the innermost annulus, where the best fit
temperature of 2.54$_{-0.57}^{+1.02}$ is marginally lower (but still
consistent within $1\sigma$ errors). We conclude that the quasar
contribution does not significantly impact the deprojected fit
parameters for the cluster observation. 

Given the flux for the models in Table~\ref{table-2b} and applying the
appropriate K-correction,
we find a total cluster luminosity L$_{0.5-2~\rm keV} = 4.6 \pm0.2 \times
10^{44}$~erg~s$^{-1}$.

\subsection{Cluster Mass}

\begin{figure*}
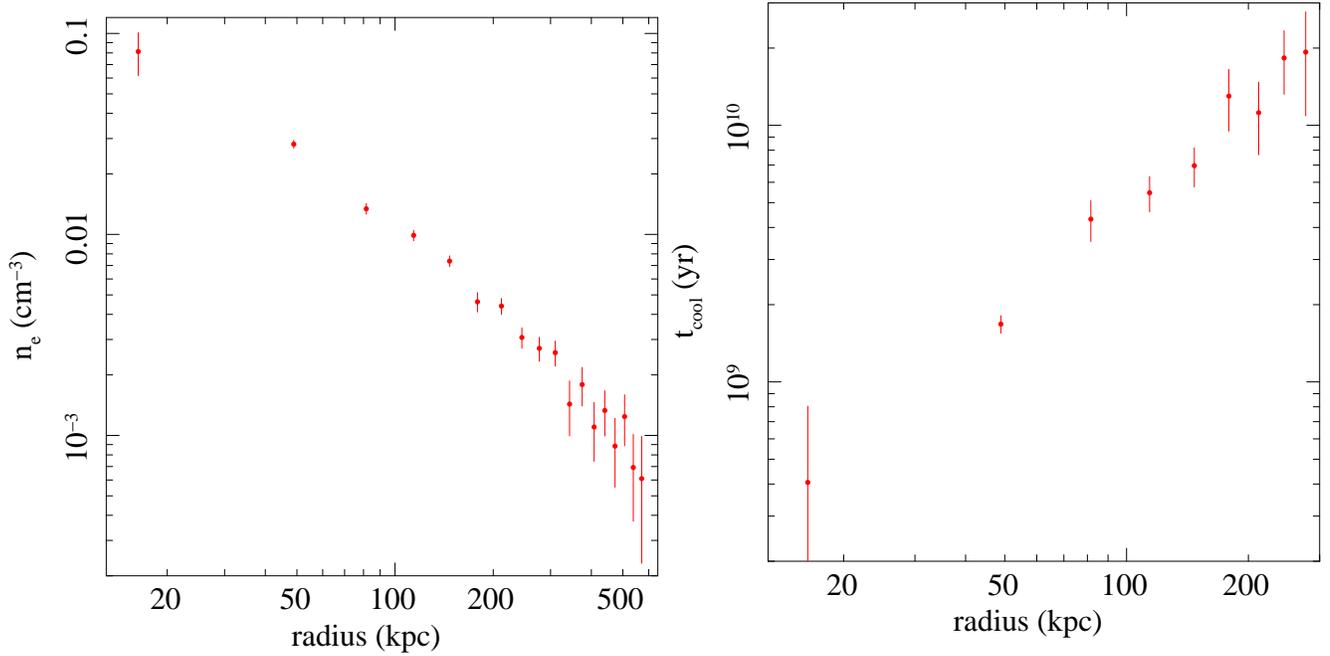

\includegraphics[angle=-90,scale=0.5]{f6a.eps}
\includegraphics[angle=-90,scale=0.5]{f6b.eps}
\caption{\small Density and cooling time profiles determined from the
  observed X-ray surface brightness and deprojected temperature
  profile, using the method of Allen et al. (2008). Error bars are
  68\% uncertainties. The results for the innermost bin include a
  conservative systematic allowance for uncertainties associated with
  modeling the central AGN emission.}
\label{density}
\end{figure*}

We have measured gas mass and total mass profiles for the cluster
using the Monte Carlo method of \cite{allen2008}. This analysis is
also used to determine gas density and cooling time profiles for the
cluster, which are shown in Figure~\ref{density}. The density results
are consistent with those obtained directly from the spectral fitting.

Our mass analysis uses a parameterized \cite{navarro1995,navarro1997}
(NFW) model, which is fitted directly to the observed cluster surface
brightness profile and deprojected temperature profile \citep[see Eq.1
and the description of the method in][]{allen2008}.  The best-fit NFW
model has a concentration parameter $c=7.4^{+2.8}_{-2.3}$, scale
radius $r_s=120^{+70}_{-40}$~kpc, and equivalent velocity dispersion
$\sigma =780^{+90}_{-60}$~km~s$^{-1}$, with  $\chi^2 = 7.9$ for
five degrees of freedom. For these parameters we calculate the radius
at which the mean enclosed mass density is 2500 times the critical
density of the Universe at the redshift of the cluster, $r_{2500}=
283^{+18}_{-13}$~kpc. This results agrees well with the results of
\cite{allen2008} based on the earlier {\it Chandra} observation.
However, the statistical uncertainties on $r_{2500}$ are improved by a
factor of $\sim 4$.

The total mass within $r_{2500}$ is 
M$_{2500} = 1.02^{+0.21}_{-0.14} \times 10^{14}$M$_{\odot}$. (Our 
68 per cent error bars on $\rm M_{2500}$ also 
account for the uncertainty in $r_{2500}$.) 
The measured gas mass fraction with this radius, $f_{gas} (r_{2500}) =
0.129^{+0.015}_{-0.016}$, is consistent with the value
determined by \cite{allen2008} from the earlier, shorter observation,  
but with significantly reduced 
statistical uncertainties.


\begin{table*}
{\scriptsize
\noindent
\caption[]{\label{table-2b} 
Best Fit Model Parameters}
\begin{center}
\begin{tabular}{rcrrcccccccc}
\hline\hline
\\
R$^a$ [arcsec] & Range [arcsec] & Total Counts$^b$  & Net Counts & kT [keV] & Norm$^c$ [1e-3] & CSTAT (dof=3550) & $n_e$[1e-2 cm$^{-3}$] \\
\\
\hline\hline

\\
 3.375  &  2.75-4.00 &603.0$\pm24.6$ &  592.6$\pm24.8$  & 3.11$_{-0.64}^{+0.91}$  &  37.1149$_{-5.6241}^{+6.0039}$ & 2922.9 & 5.91$^{+0.48}_{-0.45}$ \\
 4.875  &  4.00- 5.75 & 773.0$\pm 27.8$ & 751.8$\pm 28.2$ &   5.97$_{-1.25}^{+1.61}$  &    13.4383$_{-1.6035}^{+1.7655}$ & 3064.0 & 3.62$^{+0.22}_{-0.20}$ \\
 6.5  & 5.75-7.25 & 538.0$\pm 23.2$ & 513.9$\pm23.7$   & 4.81$_{-1.19}^{+1.61}$  &    7.4397$_{-1.1136}^{+1.1952}$ & 2976.1 & 2.65$^{+0.21}_{-0.20}$ \\
 8.75  & 7.25-10.25 & 892.0$\pm 29.9$ & 827.1$\pm 30.9$   &  7.11$_{-1.76}^{+2.43}$  &    2.6737$_{-0.2533}^{+0.3014}$ & 3231.4& 1.59$^{+0.09}_{-0.08}$ \\
 12.75  & 10.25-15.25& 1293.0$\pm 36.0$ & 1135.2$\pm 38.1$ &   7.77$_{-1.92}^{+2.93}$  &     1.0627$_{-0.0857}^{+0.0996}$ & 3388.9 & 1.00$^{+0.05}_{-0.04}$ \\
 18.25  & 15.25-21.25& 1306.0$\pm 36.1$ & 1034.8$\pm39.7$ &   6.95$_{-1.44}^{+2.88}$  &    0.4663$_{-0.0371}^{+0.0422}$ & 3460.2& 0.66$^{+0.03}_{-0.03}$\\
 25.625  & 21.25-30.00& 1452.0$\pm 38.1$ & 896.4$\pm 44.8$ &    5.03$_{-0.65}^{+0.67}$  &      0.2781$_{-0.0156}^{+0.0153}$ & 3533.8 & 0.51$^{+0.01}_{-0.01}$ \\
\\
\hline\hline
\end{tabular}
\end{center}

$^a$~The assumed annuli are circular with the mean radius listed in the R column and ranges in Range column;
$^b$~Total and net counts (in 0.5-7~keV range) summed within 4 observations in each region;
$^c$~Normalization for APEC thermal model  defined as
Norm = $ {{10^{-14}} \over { 4 \pi [D_A (1+z)]^2}} \int n_e n_H dV$ with the
abundance table set to \cite{anders};
$^d$ listed uncertainties are at 68\% for one interesting parameter.

}
\end{table*}



\begin{table*}
{\scriptsize
\noindent
\caption[]{\label{table-4} 
Best Fit Parameters for the Simulated Quasar Data}
\begin{center}
\begin{tabular}{ccrcccccccc}
\hline\hline
\\
R$^a$ [arcsec] & $\Gamma$  & Norm$^b$ [1.e-7] & \\
\\
\hline\hline
\\
 3.375  & $1.58_{-0.13}^{+0.14}$ &  $6.9_{-0.8}^{+0.9}$ \\
 4.875  & $1.58_{-0.17}^{+0.20}$ &  $4.4_{-0.7}^{+0.7}$ \\
 6.5    & $1.47_{-0.24}^{+0.26}$ &  $2.5_{-0.5}^{+0.6}$ \\
 8.75   & $1.75_{-0.23}^{+0.24}$ &  $3.1_{-0.5}^{+0.6}$ \\
 12.75  & $1.46_{-0.18}^{+0.21}$ &  $3.7_{-0.6}^{+0.7}$ \\
 18.25  & $1.35_{-0.18}^{+0.21}$ &  $1.5_{-0.4}^{+0.5}$ \\
 25.625 & $1.68_{-0.35}^{+0.42}$ &  $1.1_{-0.4}^{+0.4}$ \\
\\
\hline\hline
\end{tabular}
\end{center}

$^a$~The assumed annuli are as in Table 3, e.g. 
circular with the centers listed in the R column;
$^b$~Power law model normalization in photons~cm$^{-2}$~s$^{-1}$
$^d$ listed uncertainties are at 68\% for one interesting parameter.

}
\end{table*}


\bigskip

\section{Discussion}
\label{sec:discussion}

We have presented new, deep {\it Chandra\/} observations of the high
redshift X-ray cluster associated with 3C\,186, a luminous compact
radio-loud quasar at $z=1.067$ \citep{wills1992,sdss2007}. The new
observations confirm the main results from our discovery paper
\citep{siem2005}, including the results on the global cluster
temperature and central density profile. X-ray emission from the
cluster is detected out to $\sim 3$ times larger distance from the
quasar than was the case in the first, short observation. The quasar
is located within the center of the diffuse X-ray emission and only
slightly offset (5.5$\pm{0.9}$~kpc) from the centroid of the cluster's
X-ray emission.  The high signal to noise data and a larger cluster
area uncovered in the new observations allowed for more detailed
analysis of the properties of the cluster.  Below we discuss the main
results of these new X-ray observations.

\subsection{Cooling core}

The 3C\,186 X-ray cluster shows an elongated morphology that is
detected out to $r>500$~kpc. The cluster temperature profile has the
characteristic shape of a cooling core cluster, with a sharp decline
towards the center. The cluster is relatively cool in the outer
($r>200$~kpc) regions with its temperature increasing slightly to a
peak of 7.8$^{+2.4}_{-1.9}$~keV at 100-200~kpc and then declining to
$3.11^{+0.91}_{-0.64}$~keV in the central regions (Fig.~\ref{temp}).
The electron density rises relatively smoothly from
$\sim$0.001~cm$^{-3}$ at $r\sim500$kpc to $\sim 0.1$~cm$^{-3}$, as one
moves inward from r=500kpc to the innermost resolved regions
(Fig.~\ref{density}). The sharp drop in entropy in the inner regions
is also typical for cool core clusters.

The cooling time profile is shown in Figure~\ref{density}.  We measure
cooling times at radii of 50~kpc and 25~kpc of $1.7\pm0.2 \times
10^9$~years and $7.5\pm2.6 \times 10^8$~years, respectively.  We also
measure nominal mass cooling rate (in the absence of cooling) of
$400\pm190$\,M$_{\odot}$year$^{-1}$ within the central 100kpc.  The
cluster surrounding 3C\, 186 clearly possesses a very strong cooling
core.

The measured core radius of $\sim28.6\pm2.5$~kpc is small in
  comparison to typical core radii of nearby clusters. However,
similarly small core radii have also been observed in other
lower-redshift clusters with strongly cooling cores
\citep[e.g.][]{allen2001,schmidt2001,allen2002}.

Observationally X-ray clusters divide into two classes: cool core
clusters and non-cool core clusters.  This division is based on the
peak X-ray surface brightness and a central cooling time.
\cite{burns1990} studied radio emission of central galaxies (cD) in a
sample of Abell clusters and noticed that the cDs in cool core
clusters were more likely than the ones in non-cool core clusters to
be radio loud with a high radio power.  More recent studies confirm
that the cool core clusters are more likely to host a cD with both
radio emission and H$\alpha$ emission lines
\citep[e.g.][]{sanderson2009,haarsma2009}. \cite{mittal2009} reported
that all strong cool core clusters harbor a central radio source,
while 67\% of the weak cool cores do, and 45\% of the non-cool cores
do.

3C\,186 cluster hosts a luminous quasar with broad lines and a compact
radio structure fully contained within the host galaxy.  3C~186 is
also a high redshift ($z=1.06$) cluster with a strong cooling core.
This is contradictory to the suggestion by \cite{vikhlinin2007}, based
on a {\it Chandra} sample of X-ray clusters that there are no cool
core clusters at $z>0.5$.  On the other hand more recent work
\citep{santos2008,alshino2010} indicates that the fraction of weak to
moderate cool cores remains the same at high redshift and only the
fraction of strong cool cores drops significantly.

Is 3C\,186 unique?  It is interesting to note that a fraction of AGN
in clusters increases with redshift \citep{martini2009}, so high
redshift clusters are more likely to host an AGN. An X-ray emission
associated with AGN can confuse detection of a cooling core in a
cluster at high redshift \citep{branchesi2007}. Cool core clusters
also have smaller cooling radii and their detection require high
resolution X-ray observations.  Therefore, samples of X-ray clusters
used in studies of cool core evolution with redshift may be missing
clusters associated with a strong AGN. We note that H1821+643 cluster
at z=0.3 also has a strong cooling core that was observed by {\it
  Chandra} \citep{russell2009}. Further studies of higher redshift
clusters associated with AGN are needed in order to understand the
evolution of cluster cooling with redshift.

\subsection{Supermassive Black Hole Powering 3C\,186 Quasar}
\label{sec:bhmass}

The black hole mass estimate for the 3C\,186 quasar from measurements
of its broad emission lines is equal to $3\times 10^9$M$_{\odot}$
\citep{siem2005}. The corresponding Eddington luminosity is equal to
$L_{Edd} = 4 \times 10^{47}$~erg~s$^{-1}$.  The quasar optical-UV
luminosity, based on the spectral energy distribution given in
\cite{siem2008}, is equal to $L_{UV} \sim 6\times
10^{46}$~erg~s$^{-1}$. Using a bolometric correction that ranges
between $\sim 5-10$ \citep{elvis1994} we estimate the 3C\,186
bolometric luminosity to be of the order of $L_{bol} \sim
10^{47}$~erg~s$^{-1}$ with the required accretion rate of 0.25$\dot
M_{Edd}$ critical rate.

The growth of this supermassive black hole might be closely related to
the mass deposition from the cluster.  If the cooling rate is $\sim
470$~M$_{\odot}$~year$^{-1}$ then only a small fraction, $<0.5\%$, of
the cooling gas is needed to grow a $10^9$M$_{\odot}$ black hole
within the cooling time of the cluster's core. However, the mechanism
of transporting this gas to the close vicinity of a central black hole
is unclear.

\subsection{Intermittent Radio Source}

3C\,186 radio source belongs to a class of young CSS radio sources
\citep[see][for review]{odea1998}. \cite{murgia1999} measured a
synchrotron age of $\sim 5\times 10^5$ years for the entire radio
structure. 3C\,186 has a double radio morphology with a one-sided jet.
The double radio source has a total length equal to 1.8$\arcsec$ (see
\cite{siem2005} and references therein) corresponding to 15 kpc
projected size. A deprojected size is at least 30~kpc for $<30$
degrees angle to our line of sight (based on the one-sided VLA jet),
but probably not much larger (e.g., 100 kpc, or radius of 50 kpc, for
$\sim 9$ degrees angle which is too small).  Therefore we conclude
that the 3C\,186 radio source is contained within the host galaxy.

Studies of compact radio sources suggest that they might have
repetitive outbursts on short timescales of $\sim 10^3-10^5$~years
\citep{baum1990,begelman1997,owsianik1998,czerny2009}. In the case of
the shortest timescales ($\sim 10^3$~years) the radio source does not
have enough energy to grow beyond the host galaxy and it starts to
recollapse within the host glaxy ISM.  If there have been previous
outbursts of the radio activity in 3C\,186 on the timescales longer
than $3\times 10^4$~years, the radio source would have been larger than
the observed CSS structure. 
Our initial studies of the VLA radio data show a possible presence of
an extended radio emission on scales of 10~arcsec \citep{siem2008}.
However, this radio emission is seen at a very low significance and
more detailed analysis of new EVLA maps obtained recently have yet to
to confirm that this structure is real.

Most searches for X-ray clusters around radio-loud active galaxies
have been focused on those with large-scale radio structures.  Such
radio structures are old, triggered a long time ago ($> 10^7$ years)
and therefore have been interacting with the cluster environment for a
long time. In nearby clusters, long term ($\sim 10^8$ years)
intermittent radio activity of the central AGN is often imprinted into
the X-ray morphology of the cluster in the form of bubbles, ripples or
discontinuities in the surface brightness indicative of shocks
\citep[see][]{mcnamara2007}. GPS and CSS radio quasars are young
\citep[$<10^5$~years][for review]{murgia1999,odea1998} and have not
developed large scale radio structures. These sources, if found in
clusters, can potentially test the cluster heating process and the
significance of the luminous quasar in the evolution of the cluster.

\subsection{Quasars and X-ray Clusters}
\label{sec:quasars}

The majority of nearby clusters host a low power radio source with FRI
radio morphology that have buoyantly rising bubbles filled with radio
plasma. There are, however, a few examples of X-ray clusters
associated with quasars or powerful radio galaxies at lower redshifts
(for example Cygnus~A, 3C~295 \cite{allen2001},
IRAS~09104+4109~\cite{iwasawa2001}, HS1821+643~\cite{russell2009}). We
note that the FRII radio source would have pressure driven radio lobes
and jets and its X-ray morphology may be different than the one seen
in clusters with FRI radio sources.

Cygnus~A is a nearby FR~II radio source embedded in a bright X-ray
cluster \citep{car96}. The {\it Chandra\/} observations show a rich
filamentary structure associated with the evolution of the radio
source within the cluster and evidence for heating of the cluster gas
\citep{wilson2000}.  However, the Cygnus~A nucleus may not be in a
luminous quasar phase although its radio power is high $L_R\sim
10^{45}$~erg~s$^{-1}$ \citep{young2002,ste2008}. The nucleus is
highly absorbed, i.e. $N_H \sim 2 \times 10^{23}$~cm$^{-2}$ as measured
by \cite{young2002}, and correcting for the absorption gives the hard
X-ray luminosity of 3.7$\times 10^{44}$~erg~s$^{-1}$. \cite{young2002}
estimated the optical luminosity to be consistent with Seyfert
galaxies.

Another case of a relatively bright X-ray cluster detected around a
lower redshift z=0.322 luminous quasar is HS1821+643 \citep[$L_{bol} \sim 2
\times 10^{47}$~ erg~s$^{-1}$,][]{kolman1991,Crawford:1999,russell2009}. 
This quasar has typical signatures of a quasar with broad lines and
thermal emission in optical-UV band. It hosts a 300~kpc FRI radio
source \citep{blundell2001} that might be heating the cluster medium
\citep{russell2009}. The {\it Chandra\/} observation indicates complex
interactions between the quasar and the cluster. However, the cluster
properties are typical for the cool core cluster with a short central
cooling time and the quasar does not appear to significantly impact
the large scale cluster environment.

\subsection{Cluster Heating}
\label{sec:heating}

3C\,186 is the first high redshift X-ray cluster known to host a
luminous quasar and a compact radio source.  The cluster X-ray
morphology indicates that the cluster is well formed and has a cool
core with a short central cooling time. The radio source can potentially
supply the energy required to stabilize the cluster core against catastrophic 
cooling, as it expands into
the cluster medium.

In \cite{siem2005} we estimated the power of the radio jet using the
\cite{willott99} (their Eq.12) relation between radio luminosity and
jet power defined as $Q = 3\times 10^{38} L_{151}^{6/7}$~W, where
$L_{151}$ is in units of 10$^{28}$W~Hz$^{-1}$~sr$^{-1}$.  Assuming the
3C\,186 151~MHz flux density of 15.59~Jy \citep{hales1993} which
accounts for the total radio source emission (the radio core is
absorbed at this frequency) we find $L_{151} = 7.5 \times
10^{27}$~W~Hz$^{-1}$~sr$^{-1}$ and then from the above equation we
obtain the jet power of $L_{jet} = 2.4 \times 10^{45}$erg~s$^{-1}$.
There is significant scatter in the \cite{willott99} relation,
therefore this is an order of magnitude estimate.

The pressure in the radio lobes based on the radio flux density
measurements and the equipartition assumption is $\sim
10^{-8}$~erg~cm$^{-3}$ \citep{siem2005} and it exceeds the thermal
pressure of the cluster\footnote{Note \cite{siem2005} gives a factor
  of 10 lower value which is a mistake.} $\sim 4 \times
10^{-10}$~erg~cm$^{-3}$ (for $\rm kT=3.1$~keV and $n=0.08$~cm$^{-3}$).
The overpressured radio source should drive a strong shock into the
cluster medium and its expansion is not adiabatic.  We can estimate a
lower limit on the jet power from the equipartition measurements.
Using the radio lobes volume of $\sim 10^{66}$cm$^3$~\citep{siem2005}
the minimum pressure gives a lower estimate of the instantaneous jet
power, e.g. $\sim 10^{58}$~erg or $\sim 6 \times10^{44}$~erg~s$^{-1}$
for the age of the radio source of $5 \times 10^5$ years.

We note that the bolometric luminosity of 3C\,186 is equal to
$L_{bol} \sim 10^{47}$~erg~s$^{-1}$ (see Sec.\ref{sec:bhmass}) and the
3C\,186 radiative power exceeds the jet kinetic power by at least a
factor of 10 .  This is unusual for a cluster-center radio source and
suggests that the so-called 'quasar mode' may be more important than
the 'radio mode' for heating the 3C\,186 cluster.

The process of transferring accretion energy into the cluster thermal
energy is unclear. In the 'radio mode', the jet carries the energy
from the black hole and deposits it into the cluster gas, e.g.  via
shocks. In the 'quasar mode', the radiation should be a dominant
carrier of the accretion energy.  \cite{king2009} argues that for
quasars with a black hole mass exceeding about 10$^9$~M$_{\odot}$,
radiation can be very efficient in initiating strong outflows that
cause radiative shocks and result in cluster heating. On the other
hand, quasar radiation energy can also be transferred directly to the
cluster gas via Compton scattering.

We calculate the energy required to prevent significant cooling of the
cluster core in 3C\,186 following \cite{king2009}. The mass of the gas
within the cooling radius of 45~kpc assuming the central density of
0.08~cm$^{-3}$ is equal to $\rm M_{core} = 3.3 \times
10^{11}$~M$_{\odot}$. The amount of energy required to heat this gas
is of the order of 1~keV per baryon, e.g. $ E_{heat} \sim (1 {\rm
  keV/1 GeV})\,\rm M_{core} c^2 \sim 10^{-6} \rm M_{core} c^2 \sim
6\times 10^{59}$~erg. The observed quasar luminosity of $\sim
10^{47}$~erg~s$^{-1}$ provides enough energy to heat the cluster core
within about 2$\times 10^{5}$~years if the heating process were 100
percent efficient.  However, the cooling time of the core is much
longer, $\sim 7\times 10^8$~years, and only a very small fraction of
the observed luminosity is needed to support the cluster heating
within that time.

We estimate the efficiency of transferring the radiation energy into
the cluster.  The opacity of the cluster gas in the core to Compton
scattering is equal to $\tau = n_e r_0 \sigma_T \sim 0.008$, assuming
the average particle density of $n_e = 0.075~{\rm cm}^{-3}$,
$r_0=45$~kpc core radius, and $\sigma_T$ is the Thomson cross section.
This means that about 0.8$\%$ of the quasar photons will interact with
cluster gas and may heat up the cluster.  We note also that only the
photons with energies higher than $\sim1$ keV will be able to
heat\footnote{We note that the UV photons will also cool the cluster
  gas, however the total energy contained in the quasar spectrum above
  $> 1 keV$ exceeds the energy in the UV band.  The heating-cooling
  balance of the cluster gas has to be modeled properly to understand
  the heating efficiency. } the cluster gas to the required
temperatures which reduces the amount of available photons by a factor
of 10. We also assumed spherical symmetry and if we include a covering
factor of 30\% that accounts for cold gas in the central regions (e.g.
torus), the total number of photons decreases to about 0.3\%. This
results in the available instantenous power of $0.003 \times
10^{46}$~erg~s$^{-1}$ which is equivalent to a total energy of $\sim 7
\times 10^{59}$~erg available to balance the cluster cooling within
7$\times 10^8$~years. This is within an order of magnitude of the
required energy to balance the cluster cooling.  However, this also
requires that the quasar is powered by accreting at 0.25$\dot M_{Edd}$
rate, (e.g. $\sim 18 \dot M_{\odot}$~year$^{-1}$ assuming standard
10\% accretion efficiency), continuously during the cooling time of
the cluster, which results in the final required amount of fuel
exceeding by a factor of 4 the current black hole mass.  Although such
an amount of fuel is only a fraction of the available mass in the
cluster core, it is unclear how it can be funneled down to the center
within close vicinity of a black hole.

\cite{king2009} suggested that the radiation pressure in the luminous
quasars will generate a wind, so the radiation energy is transferred
to the kinetic energy of the wind and then to the cluster gas via
radiative shocks. This mechanism also requires a large amount of fuel
supply.

We should comment here that the young CSS radio sources, such as
3C\,186, might be intermittent on short timescales $ \sim 10^5$ years
if the accretion rates are close to the Eddington value.
\citep{czerny2009}. The source experiences transitions between (1) a
high super-Eddington state characterize by a luminous accretion disk
and a powerful jet; and (2) a quiescent state with the sub-Eddington
luminosity of the disk and no jet. In this scenario the
super-Eddington state should last longer than $\sim 10^3$~years for a
radio source to grow beyond the host galaxy, as the expansion of the
radio source within the host galaxy takes more than $\sim 10^6$~years.
During that initial time the luminous quasar can provide enough
radiative power to heat up a small cluster core. At later times when
the radio source grows beyond the host galaxy and expands within the
cluster environment the mechanical energy is directly used to prevent
cluster cooling.

\subsection{Clusters at High Redshift and Cosmology}
\label{sec:mass}

X-ray observations of galaxy clusters at high redshift can provide
strong constraints on cosmological parameters (see
\cite{vikhlinin2010} for a recent review). In hierarchical models of
cluster formation, the high-mass end of the mass function is the most
sensitive to the linear growth of the fluctuations \citep{linder2003}.
Thus the evolution of the number density of massive clusters that
traces the growth of the density fluctuation can be used to constrain
the dark energy equation of state parameter {\it w} \citep{haiman2001,
  alexy2009, mantz2010}. The gas mass fraction of clusters can also be 
used to test the cosmological parameters \citep{allen2008, allen2004}
and provide independent constraints on {\it w}.

\cite{allen2008} measured $f_{gas} (r_{2500})$ for 42 clusters at $0<z<1.1$ 
and obtained an average value of $f_{gas} (r_{2500}) = 0.1104\pm0.0003$. 
Their study included the previous, short observation of the
3C\,186 cluster (15ks good time), 
which was the highest redshift cluster in their sample.
The measured gas mass fraction for this target by \cite{allen2008} was 
$f_{gas} (r_{2500}) = 0.1340\pm0.0777$. Our new
measurement of  $f_{gas} (r_{2500}) = 0.129^{+0.015}_{-0.016}$ is
consistent with the previous value, but improves the statistical 
uncertainties by a factor of $\sim 4-5$. 
Our new $f_{gas}$ measurement for the 3C\,186 cluster is also consistent, 
within measurement errors, with the mean value determined at lower 
redshifts, arguing against any strong evolution of the 
$f_{gas} (r_{2500})$ value for massive, relaxed clusters over
the redshift range $0<z<1.1$.

A relatively small number of X-ray clusters at $z>1$ have been
observed so far (The BAX\footnote{http://bax.ast.obs-mip.fr/}
\citep{sadat2004} data base lists 17 $z>1$ clusters to date). New
methods to search for high-redshift X-ray clusters are providing new
discoveries of massive clusters,
including the recent discovery of the most distant X-ray cluster at
$z=1.99$ \citep{andreon2009}. The 3C\,186 cluster is so far the only
spectroscopically confirmed cooling core cluster at such high redshifts, 
and the only one known to host a luminous quasar at the center.
The cluster global temperature, $kT = 5.74$~keV is not extreme 
when compared to the other massive clusters at $z>1$,
although the X-ray luminosity,
L$_{0.5-2 \rm keV} = 6.4 \times 10^{44}$~erg~s$^{-1}$, 
exceeds that of the next most luminous
cluster detected so far, {\it XMMU~J2235.3-2557} \citep{rosati2009} at
$z=1.39$. The measured 3C\,186 cluster metallicity lies slightly above 
the mean trend with redshift discussed by
\cite{balestra2007}. More detailed evolutionary studies should soon 
become possible with
improved high redshift cluster samples
from {\it Chandra} and XMM-{\it Newton}.

\section{Summary and Conclusions}
\label{sec:summary}

We have presented results from a deep (200ks) {\it Chandra\/} image of 
the hot ($kT=5.6\pm0.3$keV), X-ray luminous galaxy 
cluster surrounding the powerful quasar 3C186 at a redshift $z=1.067$.  
The spatially resolved temperature profile, entropy, density and 
cooling time profiles all confirm 3C\,186 as a cooling core cluster. 
The measured gas mass fraction of $f_{gas} = 0.129^{+0.015}_{-0.016}$
at $r_{2500}$ is consistent with measurements for lower redshift 
systems. This argues against strong evolution in $f_{gas}(z)$ at 
$r_{2500}$ for massive, relaxed systems. Cooling gas in the cluster core 
can in principle support the growth of a supermassive black hole and
power the luminous quasar.  The radiative power of the quasar exceeds the
kinematic power, suggesting that radiative heating my be important at
intermittent intervals in cluster cores.

\section*{Acknowledgments}

We thank the anonymous referee for comments that greatly improved the
manuscript. AS thanks Mitch Begelman and Anna Wolter for comments.  We
thank Agnieszka Siemiginowska for English improvement to the text.
This research is funded in part by NASA contract NAS8-39073.  Partial
support for this work was provided by the {\it Chandra\/} grants
GO2-3148A, GO5-6113X and GO8-9125A-R. Basic research in radio
astronomy at the NRL is supported by 6.1 Base funding.

{}

\appendix

\medskip

\centerline{\bf A1: Modeling Background Spectra}

\medskip

\begin{figure}
\includegraphics[width=\columnwidth]{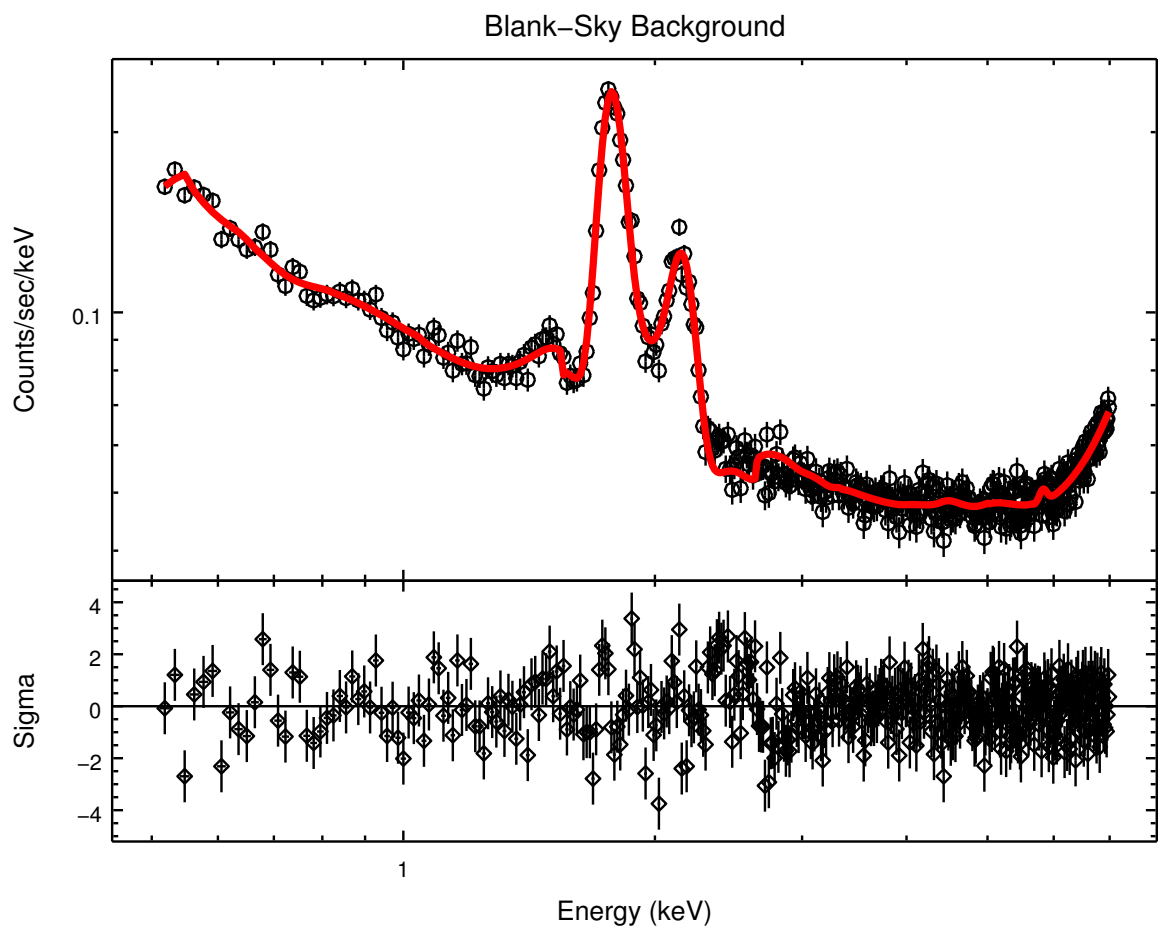}
\caption{\small {\bf Top panel:} Blank-sky background data fit with the complex model: 
polynomial + 9 gauss lines. {\bf Lower panel:} Residuals.}
\label{blanksky}
\end{figure}

We used {\it Sherpa} to fit simultaneously source and background data
in the spectral analysis of the cluster.  A ``blank-sky'' background
file {\tt acis7sD2005-09-01bkgrnd\_ctiN0001.fits} provided by the {\it
  Chandra} X-ray Center was used to define a background model. First
we filtered and reprojected X-ray events contained in this background
file following the CIAO Thread {\it The ACIS "Blank-Sky" Background
  Files}\footnote{http://cxc.harvard.edu/ciao/threads/acisbackground/index.py.html}.
Next we extracted a spectrum from a box region ({\tt
  rotbox(4385.5,4112.5,980,990,0)}, the first two values show a center
of the box, the next ones the size of the box in ACIS pixels, and a
rotation angle) and corresponding instrument response files using {\tt
  specextract} tool in CIAO. We fit this spectrum in {\it Sherpa}
using a combination of an 8th-order polynomial and 9 gaussian lines
getting the best-fit model with the statistics equal to {\tt
  cstat}=540.8 (443 d.o.f). Figure~\ref{blanksky} shows the resulting
fit and residuals.

\begin{figure}
\includegraphics[width=\columnwidth]{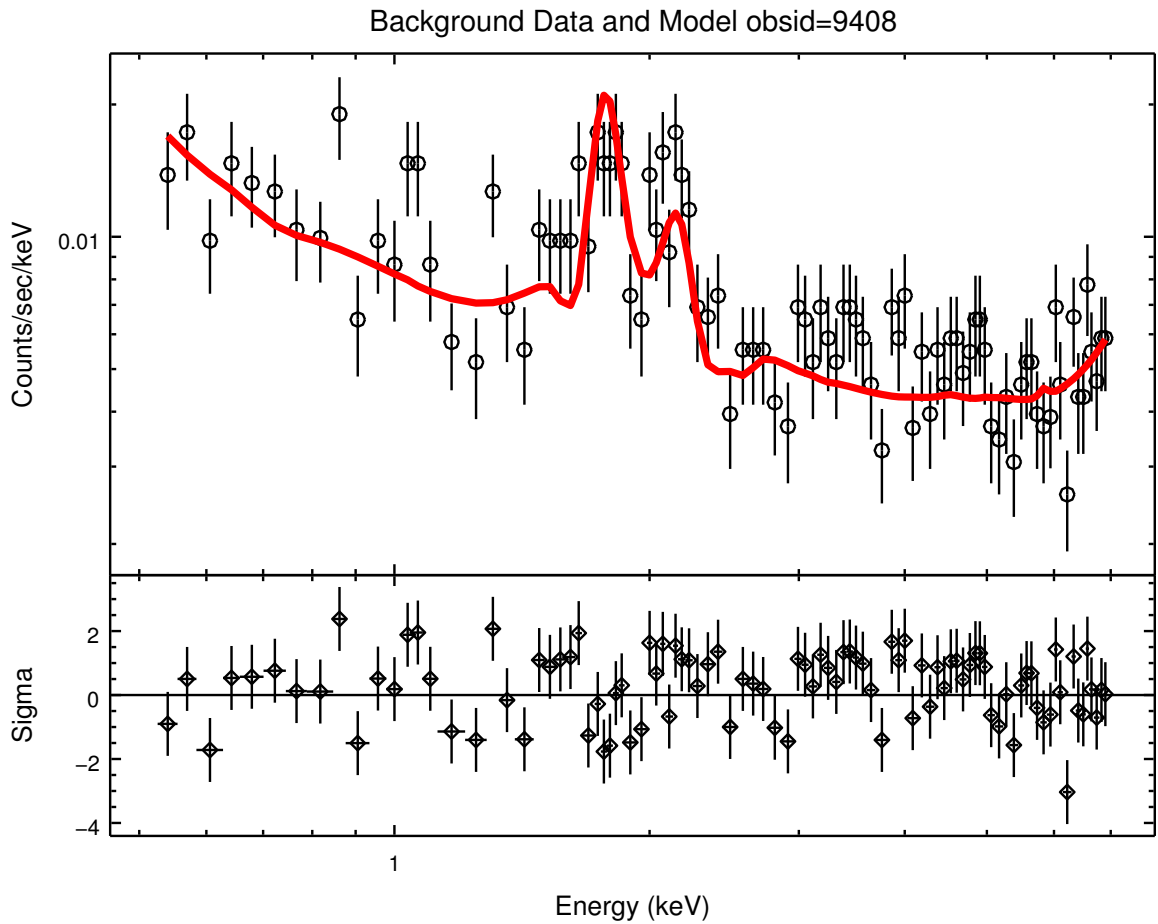}
\caption{\small The background spectrum for obsid 9408 fit with the
  same shape background model as the blank-sky data shown in
  Fig.~\ref{blanksky}. Top panel shows the data marked with points and a
  model drawn with a red solid line. The bottom panel shows the
  residuals. The data were grouped by 15 counts per bin for the
  visualization only.}
\label{9408}
\end{figure}

We next checked how well this empirical model fits the background data
in the observations. We applied the background model to the background
spectra in each obsid and fit only the model normalization. The
resulting fit statistics is equal to {\tt cstat=1900.4} (1772 d.o.f.)
for the simultaneous background fit to all four observations.
Figure~\ref{9408} shows the fit result for the obsid 9408 with the
highest background counts in the spectrum.

In the final simultaneous source and background fitting of the cluster
spectra the background normalization was varied and appropriate
background model predicted counts were included in the total model
predicted counts. An additional constant scaling accounts for a
difference in exposure times and areas of the background and source
regions and it is automatically applied by {\it Sherpa} during the fit.


\label{lastpage}

\end{document}